\documentclass[12pt]{article}
\usepackage{amsmath}
\usepackage{graphicx}
\usepackage{cite}
\usepackage[latin5]{inputenc}
\usepackage{subfigure}
\usepackage{caption}
\usepackage{amssymb}
\usepackage{float}
\restylefloat{table}
\DeclareMathOperator{\sech}{sech}
\DeclareMathOperator{\csch}{csch}
\begin{document}
\title{\textbf{Exact Solutions of Space-time Fractional EW and modified EW equations}}
\author{Alper Korkmaz{\thanks{ Corresponding Author: akorkmaz@karatekin.edu.tr}}\\
Department of Mathematics, \\
Çankırı Karatekin University, Çankırı, TURKEY\\}
\maketitle
\begin{abstract}
The bright soliton solutions and singular solutions are constructed for space-time fractional EW and modified EW equations. Both equations are reduced to ordinary differential equations by the use of fractional complex transform and properties of modified Riemann-Liouville derivative. Then, implementation of ansatz method the solutions are constructed. 
\end{abstract}

\section{Introduction}
Several decades ago, more generalized forms of differential equations are described as fractional differential equations. Various phenomena in many natural and social sciences fields like engineering, geology, economics, meteorology, chemistry and physics are modeled by those equations\cite{podlubny,guo}. The descriptions of diffusion, diffusive convection, Fokker-Plank type, evolution, and other differential equations are expanded by using fractional derivatives. Some well known fractional partial differential equations (FPDE) in literature can be listed as diffusion equation, nonlinear Schrödinger equation, Ginzburg-Landau equation, Landau-Lifshitz, Boussinesq equations, etc.\cite{guo}.

\noindent
Even though there exist general methods for solutions of linear partial differential equations, the class of nonlinear partial differential equations have usually exact solutions. Sometimes it is also possible to obtain soliton-type solitary wave solutions, which behaves like particles, that is, maintains its shape with constant speed and preserves its shape after collision with another soliton, for partial differential equations. The famous nonlinear partial differential equations having soliton solutions in literature are KdV, and Schrödinger equations. Soliton type solutions have great importance in optics, fluid dynamics, propagation of surface waves, and many other fields of physics and various engineering branches.

\noindent
The integer ordered form 
\begin{equation}
U_{t}(x,t)-U(x,t)U_{x}(x,t)-U_{xxt}(x,t)=0 \label{cEWE}
\end{equation}
was named as the Equal-width Equation (EWE) by Morrison et al.\cite{morrison} due to having traveling wave solutions containing $\sech^2{}$ function. The EWE has only lowest three polynomial conservation laws and they were determined in the same study. The single traveling wave solutions to the generalized form of the EWE are classified by implementing the complete discrimination system for polynomial\cite{fan}. Owing to having analytical solutions, the EWE also attracts many researchers studying numerical techniques for partial differential equations. So far, various numerical methods covering differential quadrature, Galerkin and meshless methods\cite{ew3}, lumped Galerkin methods based on B-splines\cite{esen,kkoc}, septic B-spline collocation\cite{kkoc2}, the method of lines based on meshless kernel \cite{dereli} have been applied to solve the EWE numerically.

\noindent
Recently, parallel to developments in symbolic computations, lots of new techniques have been proposed to solve nonlinear partial differential equations exactly. Some of those methods covering the first integral method, the sub-equation method, Kudryashov method, and ansatz methods have been applied for exact solutions for not only integer ordered and but also fractional ordered partial differential equations\cite{ozkan11,ozkan12,biswas11,biswas12,biswas13}. Some recent studies including various methods for exact solutions of fractional partial differential equations in literature can be found in\cite{ozkan1,ozkan2,ozkan3,ozkan4,ozkan5,ozkan6,biswas1,akbari}. 

This study aims to generate exact solutions for the space-time fractional equal-width equation (FEWE) and modified fractional equal-width equation (MFEWE) of the forms
\begin{equation}
\begin{aligned}
D_{t}^{\alpha}u(x,t)+\epsilon D_{x}^{\alpha}u^2(x,t)-\delta D_{xxt}^{3\alpha}u(x,t)=0 \label{FEWE}
\end{aligned}
\end{equation}
and 
\begin{equation}
\begin{aligned}
D_{t}^{\alpha}u(x,t)+\epsilon D_{x}^{\alpha}u^3(x,t)-\delta D_{xxt}^{3\alpha}u(x,t)=0 \label{MFEWE}
\end{aligned}
\end{equation}
where $\epsilon$ and $\delta$ are real parameters and the modified Riemann-Liouville derivative (MRLD) operator of order $\alpha$ for the continuous function $u:\mathbb{R} \rightarrow \mathbb{R}$ defined as
\begin{equation}
D_{x}^{\gamma}u(x)=\left\{ 
\begin{array}{lcc}
\dfrac{1}{\Gamma{(-\gamma)}}\int\limits_{0}^{x}(x-\eta)^{-\gamma-1}u(\eta )d\eta, \quad \gamma <0 \\ 
\dfrac{1}{\Gamma{(-\gamma)}} \dfrac{d}{dx}\int\limits_{0}^{x}(x-\eta)^{-\gamma}[u(\eta )-u(0)]d\eta, \quad 0<\gamma <1 \\
(u^{(p)(x)})^{(\gamma -p)}, \quad p \leq \gamma \leq p+1, \quad p\geq 1
\end{array}%
\right.  \label{derivative}
\end{equation} where the Gamma function is given as
\begin{equation}
\Gamma(\gamma )= \lim\limits_{p\rightarrow \infty}^{}\dfrac{p!p^{\gamma}}{\gamma (\gamma +1) (\gamma +2) \ldots (\gamma +p)}
\end{equation} \cite{juma}.
 
\section{The properties of the MRLD and Methodology of Solution}
Some properties of the MRLD can be listed as
\begin{equation}
\begin{aligned}
D_{x}^{\alpha} x^c&=\dfrac{\Gamma{(1+c)}}{\Gamma{(1+c-\alpha)}} \\
D_{x}^{\alpha} \{ aw(x)+bv(x) \}&= aD_{x}^{\alpha} \{ w(x) \} +bD_{x}^{\alpha} \{ v(x) \}
\end{aligned}
\end{equation} where $a,b$ are constants and $c \in \mathbb{R}$ \cite{qfrac}.

\noindent
Consider the nonlinear FPDEs of the general implicit polynomial form
\begin{equation}
F(u,D_{t}^{\alpha} u,D_{x}^{\theta} u,D_{t}^{\alpha}  D_{t}^{\alpha} u,D_{t}^{\alpha}  D_{t}^{\alpha} u, D_{t}^{\alpha}  D_{x}^{\theta} u, D_{x}^{\theta}  D_{x}^{\theta} u, \ldots )=0 , \quad 0 < \alpha , \theta <1 \label{gfpde}
\end{equation}
where $\alpha$ and $\theta$ are orders of the MRLD of the function $u=u(x,t)$. The fractional complex transform 
\begin{equation}
u(x,t)=U(\xi ), \quad \xi = \dfrac{\tilde{k}x^{\theta}}{\Gamma{(1+\theta)}}- \dfrac{\tilde{c}t^{\alpha }}{\Gamma{(1+\alpha)}} \label{fct}
\end{equation} where $\tilde{k}$ and $\tilde{c}$ are nonzero constants reduces (\ref{gfpde}) to an integer order differential equation \cite{li2010}. One should note that the chain rule can be calculated as 
\begin{equation}
\begin{aligned}
D_{t}^{\alpha} u &= \sigma_{1} \dfrac{dU}{d\xi} D_{t}^{\alpha} \xi \\
D_{x}^{\alpha} u &= \sigma_{2} \dfrac{dU}{d\xi} D_{x}^{\alpha} \xi \\
\end{aligned}\label{cr}
\end{equation}
where $\sigma_{1}$ and $\sigma_{2}$ fractional indices\cite{cr2012}. Substitution of fractional complex transform (\ref{fct}) into (\ref{gfpde}) and usage of chain rule defined (\ref{cr}) converts (\ref{gfpde}) to an ordinary differential equation of the polynomial form
\begin{equation}
G(U,\frac{dU}{d\xi},\frac{d^2U}{d\xi^2},\ldots)=0
\end{equation}
\section{Solutions for fractional EW equation}

\noindent
Consider the FEWE equation
\begin{equation}
\begin{aligned}
D_{t}^{\alpha}u(x,t)+\epsilon D_{x}^{\alpha}u^2(x,t)-\delta D_{xxt}^{3\alpha}u(x,t)=0 \label{FEWE1}\\
t>0, \quad 0<\alpha \leq 1
\end{aligned}
\end{equation}
The use of the transformation (\ref{fct}) reduces the FEWE (\ref{FEWE1}) to
\begin{equation}
-cU'+\epsilon k (U^2)'+\delta ck^2 U'''=0 \label{fewode}
\end{equation} where $k=\tilde{k}\sigma_2$ and $c=\tilde{c}\sigma_1$
\subsection{Bright soliton solution}
\begin{figure}[htp]
    \subfigure[$\alpha=0.25$]{
   \includegraphics[scale =0.4] {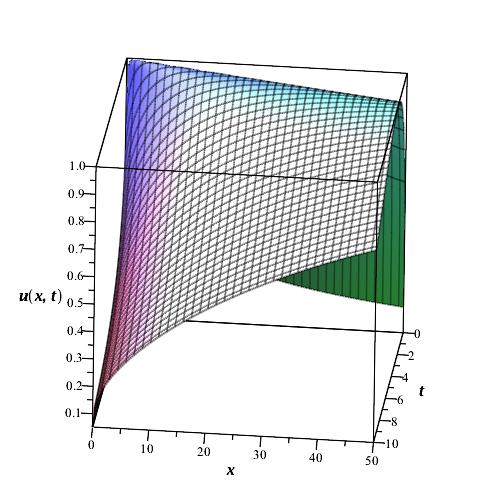}
   \label{fig:1a}
 }
 \subfigure[$\alpha=0.50$]{
   \includegraphics[scale =0.4] {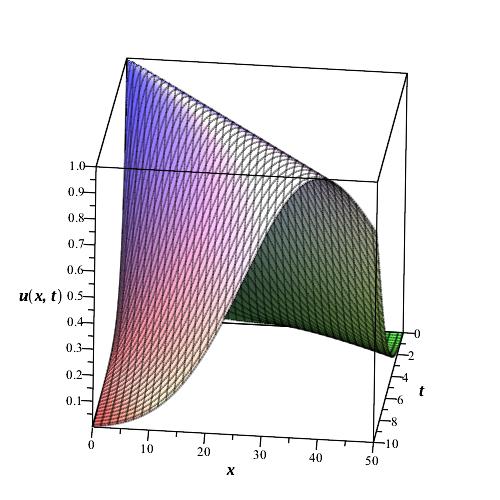}
   \label{fig:1b}
 }
  \subfigure[$\alpha=0.75$]{
   \includegraphics[scale =0.4] {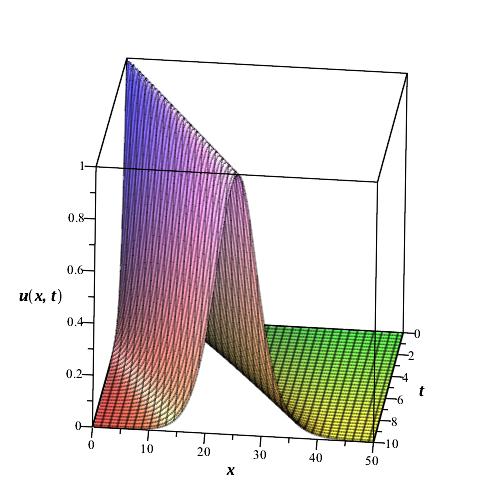}
   \label{fig:1c}
 }
  \subfigure[$\alpha=1.00$]{
   \includegraphics[scale =0.4] {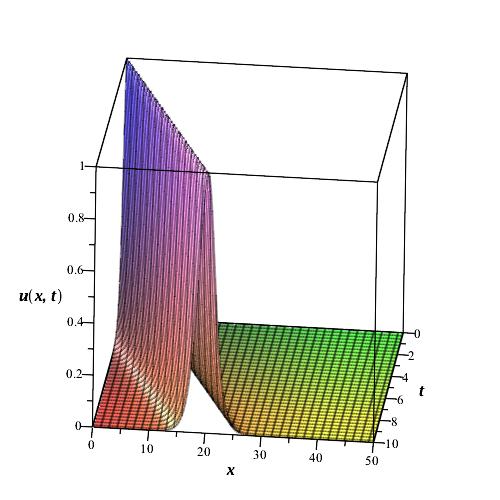}
   \label{fig:1d}
 }
 \caption{Bright soliton solutions of FEWE for various values of $\alpha$}
\end{figure}

Let $A$, $\tilde{k}$ and $\tilde{c}$ be arbitrary constants. Then, assume that 
\begin{equation}
U(\xi)=A\sech^p{\xi}, \quad \xi = \dfrac{\tilde{k}x^{\alpha}}{\Gamma{(1+\alpha)}}- \dfrac{\tilde{c}t^{\alpha }}{\Gamma{(1+\alpha)}} \label{bright}
\end{equation} 
solves Eq. (\ref{fewode}). Substituting the solution into the equation (\ref{fewode}) leads to
\begin{equation}
(-\delta\,cp{k}^{2}A-\delta\,cp^2{k}^{2}A) \sech^{p+2}{\xi}+\epsilon k A^2 \sech^{2p}{\xi} +(\delta ck^2p^2-cA)\sech^{p}{\xi}=0 \label{eq1}
\end{equation}
Equating the powers $p+2=2p$ gives $p=2$. Substituting $p=2$ into Eq.(\ref{eq1}) reduces it to 
\begin{equation}
 \left( -6\,\delta\,c{k}^{2}A+\epsilon\,k{A}^{2} \right)   
\sech^{4} \xi + \left( 4\,\delta\,c{k}^{2}A-cA
 \right) \sech^{2} \xi =0 \label{solve}
\end{equation} and solving Eq.(\ref{solve}) for nonzero $\sech^4{\xi}$ and $\sech^2{\xi}$ gives
\begin{equation}
\begin{aligned}
A&=\mp 3c\frac {\sqrt{\delta}}{\epsilon} \\
k&=\mp \frac{1}{2} \sqrt{\frac{1}{\delta}} 
\end{aligned}\label{Ak}
\end{equation}
Thus the bright soliton solution is formed as
\begin{equation}
u(x,t)=A\sech^ {2} \left ( \dfrac{\tilde{k}x^{\alpha}}{\Gamma{(1+\alpha)}}-\frac{\tilde{c}t^{\alpha}}{\Gamma{(1+\alpha)}}\right)
\end{equation}
where $A$ is given in (\ref{Ak}). The simulations of motion of bright solitons for various values of $\alpha$ are demonstrated in Fig(\ref{fig:1a}-\ref{fig:1d}) for $\delta=1$, $\epsilon=3$ and $c=1$.
\subsection{Singular Solution}

\begin{figure}[thp]
    \subfigure[$\alpha=0.25$]{
   \includegraphics[scale =0.38] {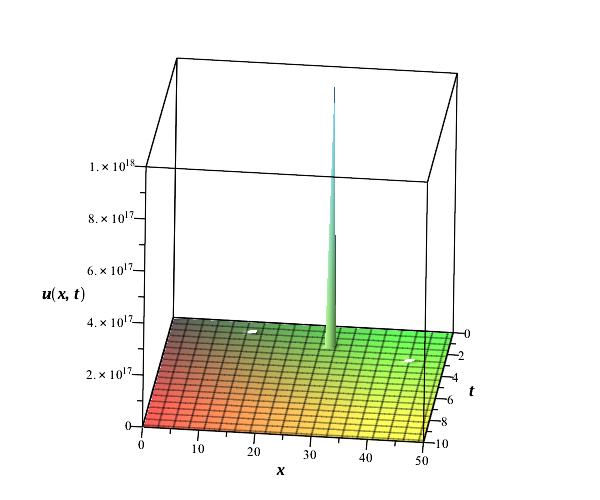}
   \label{fig:2a}
 }
 \subfigure[$\alpha=0.50$]{
   \includegraphics[scale =0.38] {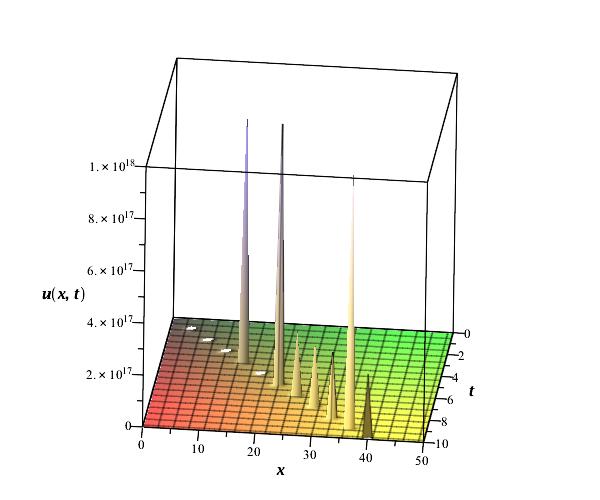}
   \label{fig:2b}
 }
  \subfigure[$\alpha=0.75$]{
   \includegraphics[scale =0.38] {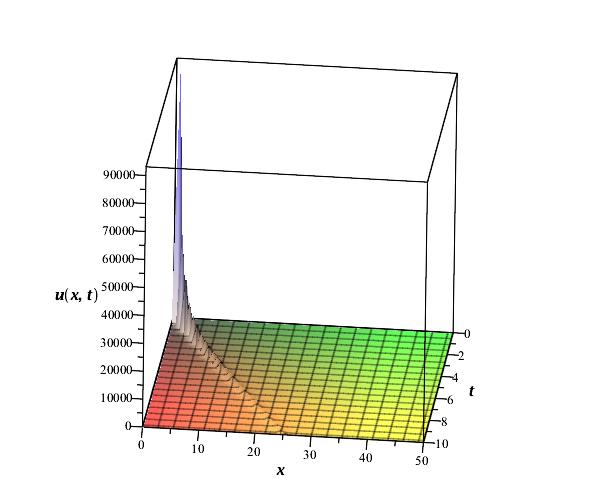}
   \label{fig:2c}
 }
  \subfigure[$\alpha=1.00$]{
   \includegraphics[scale =0.38] {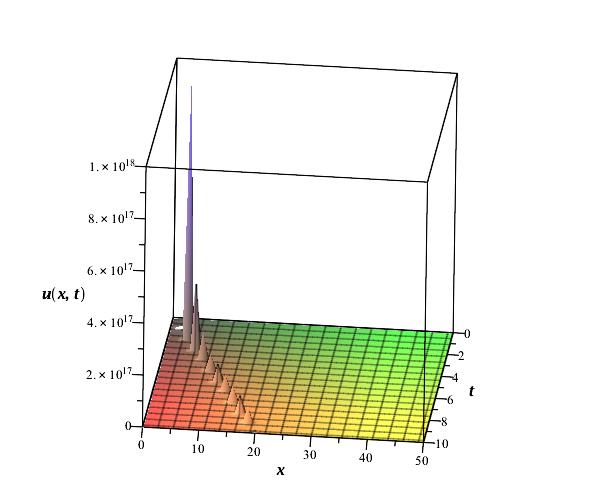}
   \label{fig:2d}
 }
 \caption{Singular solutions of FEWE for various values of $\alpha$}
\end{figure}

\noindent
Let 
\begin{equation}
U(\xi)=A\csch ^p{\xi}, \quad \xi = \dfrac{\tilde{k}x^{\alpha}}{\Gamma{(1+\alpha)}}- \dfrac{\tilde{c}t^{\alpha }}{\Gamma{(1+\alpha)}} \label{singular}
\end{equation} 
be a solution for the equation (\ref{fewode}) with $A$, $\tilde{k}$ and $\tilde{c}$ arbitrary constants. Since the solution has to satisfy the equation (\ref{fewode}), substituting it into the equation gives
\begin{equation}
 \left(\delta pc{k}^{2}A + \delta p^2c{k}^{2}A\right) \csch^{p+2}{\xi}+\left( \delta p^2c{k}^{2}A -Ac\right)\csch^{p}{\xi}+\epsilon k A^2 \csch^{2p}{\xi}=0 \label{eq3}
\end{equation}
Choosing $p+2=2p$ gives $p=2$ and reduces (\ref{eq3}) to
\begin{equation}
\left( 6\,\delta\,c{k}^{2}A+\epsilon\,k{A}^{2} \right) \csch^{4}{\xi}+\left( 4\,\delta\,c{k}^{2}A-Ac
 \right)
 \csch^{2}{\xi}=0 \label{eq4}
\end{equation}
Solution of (\ref{eq4}) for nonzero $\csch$ function give
\begin{equation}
\begin{aligned}
A&=\pm \dfrac{3c\sqrt{\delta}}{\epsilon} \\
k&=\pm \dfrac{1}{2} \sqrt{\dfrac{1}{\delta}} \label{singsol}
\end{aligned}
\end{equation}
Thus the singular solution of Eq.(\ref{fewode}) becomes
\begin{equation}
U(\xi)=A\csch ^p{\dfrac{\tilde{k}x^{\alpha}}{\Gamma{(1+\alpha)}}- \dfrac{\tilde{c}t^{\alpha }}{\Gamma{(1+\alpha)}}} \label{singular1}
\end{equation} 
where $A$ and $k$ are given (\ref{singsol}). The singular solution simulations for various $\alpha$ values and $\epsilon=3$, $\delta =c=1$ are plotted in Fig.(\ref{fig:2a}-\ref{fig:2d}).

\section{Solutions for fractional modified EW equation}
Consider space-time fractional modified equally width equation of the form
\begin{equation}
\begin{aligned}
D_{t}^{\alpha}u(x,t)+\epsilon D_{x}^{\alpha}u^3(x,t)-\delta D_{xxt}^{3\alpha}u(x,t)=0 \label{MFEWE1}
\end{aligned}
\end{equation}
The use of the transformation (\ref{fct}) converts the MFEWE (\ref{MFEWE1}) to
\begin{equation}
-cU'+\epsilon k (U^3)'+\delta ck^2 U'''=0 \label{fmewode}
\end{equation} where $k=\tilde{k}\sigma_2$ and $c=\tilde{c}\sigma_1$.
\subsection{Bright Soliton Solution}

\begin{figure}[thp]
    \subfigure[$\alpha=0.25$]{
   \includegraphics[scale =0.4] {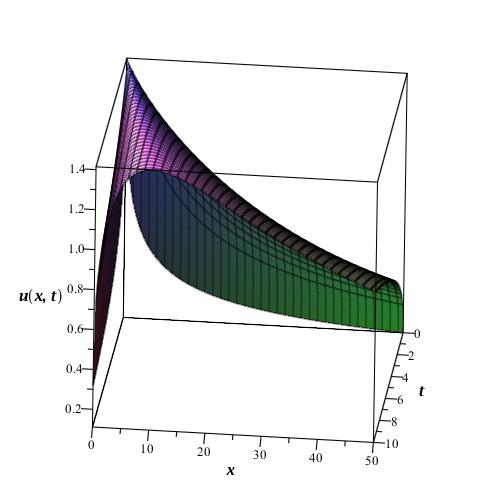}
   \label{fig:3a}
 }
 \subfigure[$\alpha=0.50$]{
   \includegraphics[scale =0.4] {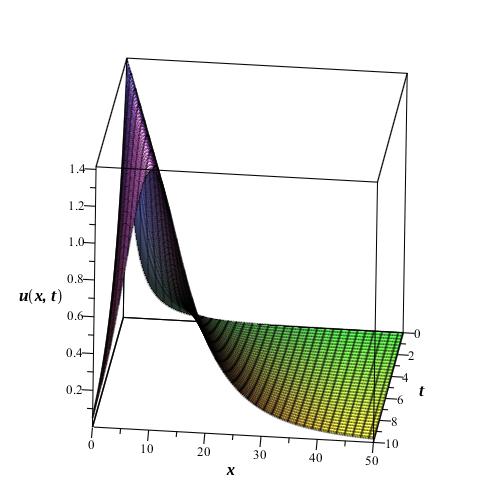}
   \label{fig:3b}
 }
  \subfigure[$\alpha=0.75$]{
   \includegraphics[scale =0.4] {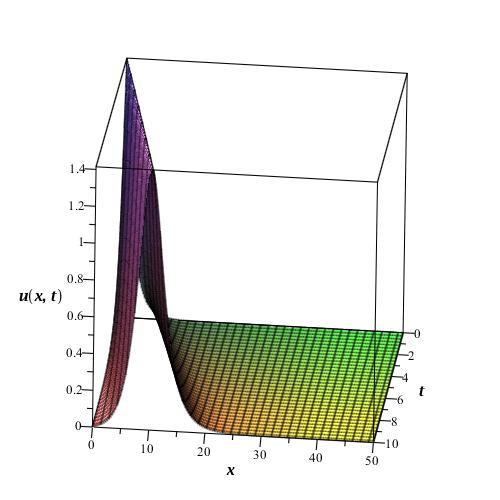}
   \label{fig:3c}
 }
  \subfigure[$\alpha=1.00$]{
   \includegraphics[scale =0.4] {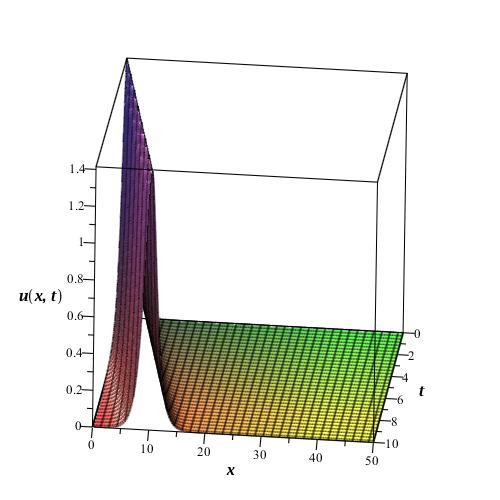}
   \label{fig:3d}
 }
 \caption{Bright soliton solutions of MFEWE for various values of $\alpha$}
\end{figure}

\noindent
Assume that the FMEWE (\ref{MFEWE1}) has a solution of the form (\ref{bright}). Substitution of (\ref{bright}) into the FMEWE (\ref{MFEWE1}) generates the equation
\begin{equation}
\left( -\delta cAk^2p-\delta ck^2Ap^2 \right) \sech^{p+2}{\xi}+ \left( \delta cAk^2p^2-cA \right) \sech^{p}{\xi}+\epsilon k A^3 \sech^{3p}{\xi}=0 \label{eqq1}
\end{equation}
Equating the powers $p+2=3p$ gives $p=1$. Substituting $p=1$ into (\ref{eqq1}) and rearrangement gives the equation
\begin{equation}
\left( \epsilon k{A}^{3}-2\delta c{k}^{2}A \right) \sech^{3}{\xi}+ \left( \delta c{k}^{2}A-cA \right) \sech{\xi}=0 \label{eqq2}
\end{equation}
Solving the coefficients of (\ref{eqq2}) for $A$ and $k$ under the assumption $\sech{\xi}\neq 0$ gives
\begin{equation}
\begin{aligned}
A&=\pm \sqrt{2} \sqrt{\pm \frac{c\sqrt{\delta}}{\epsilon}} \\
k&=\pm \frac{1}{\sqrt{\delta}}
\end{aligned}
\end{equation}
Thus, the bright soliton solution for the FMEWE is constructed as
\begin{equation}
u(x,t)=A\sech \left ( \dfrac{\tilde{k}x^{\alpha}}{\Gamma{(1+\alpha)}}-\frac{\tilde{c}t^{\alpha}}{\Gamma{(1+\alpha)}}\right)
\end{equation}
where $A$ is given as above. The simulations of bright solitons are depicted for various $\alpha$ values in Fig(\ref{fig:3a}-\ref{fig:3d}). 
 
\subsection{Singular Solution}

\begin{figure}[thp]
    \subfigure[$\alpha=0.25$]{
   \includegraphics[scale =0.38] {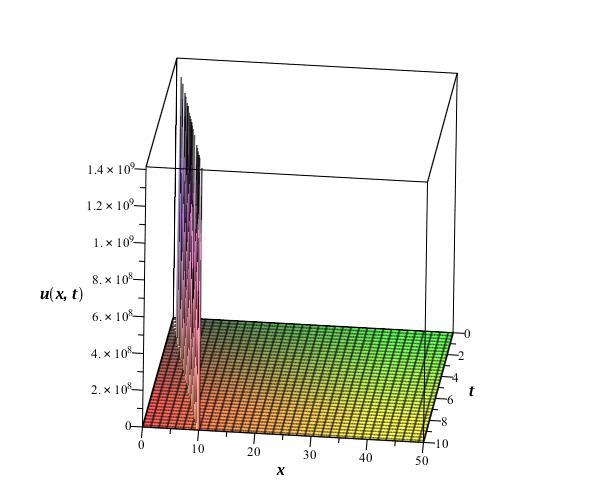}
   \label{fig:4a}
 }
 \subfigure[$\alpha=0.50$]{
   \includegraphics[scale =0.38] {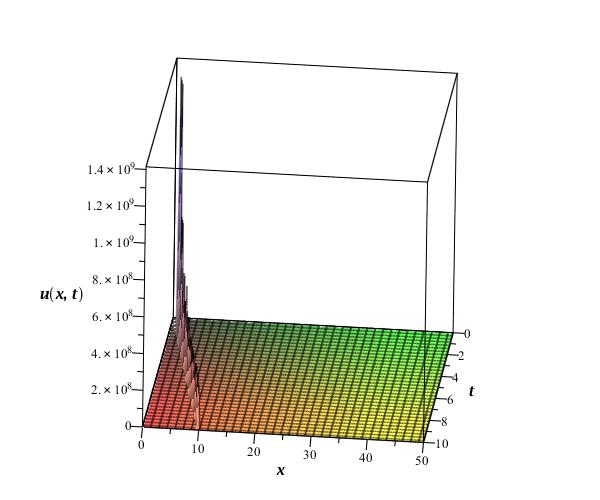}
   \label{fig:4b}
 }
  \subfigure[$\alpha=0.75$]{
   \includegraphics[scale =0.38] {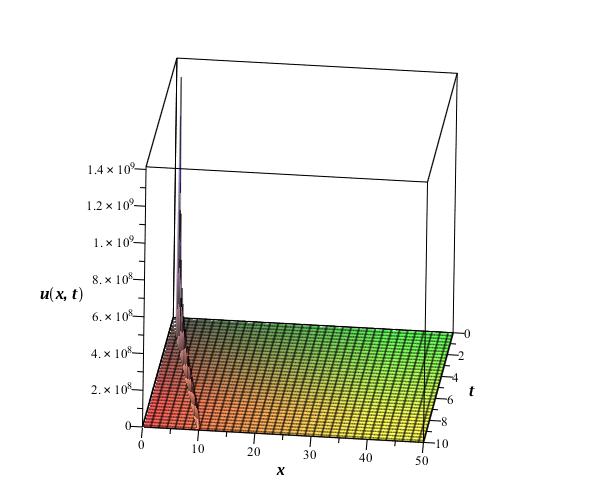}
   \label{fig:4c}
 }
  \subfigure[$\alpha=1.00$]{
   \includegraphics[scale =0.38] {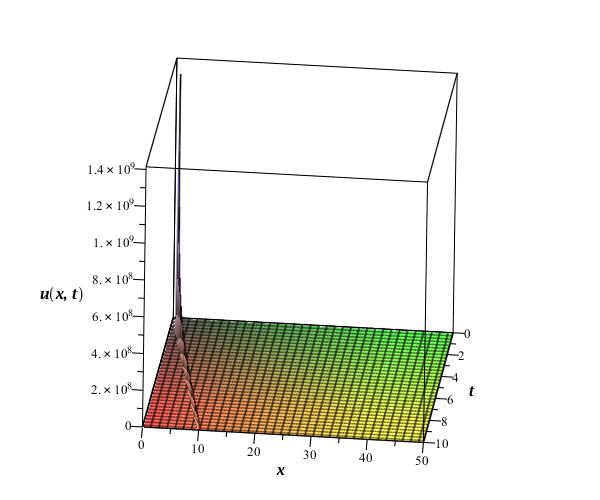}
   \label{fig:4d}
 }
 \caption{Singular solutions of FMEWE for various values of $\alpha$}
\end{figure}

\noindent
Assume that  
\begin{equation}
U(\xi)=A\csch ^p{\xi}, \quad \xi = \dfrac{\tilde{k}x^{\alpha}}{\Gamma{(1+\alpha)}}- \dfrac{\tilde{c}t^{\alpha }}{\Gamma{(1+\alpha)}} \label{singularf}
\end{equation} 
is a solution for the equation (\ref{fmewode}) with $A$, $\tilde{k}$ and $\tilde{c}$ arbitrary constants. Substitution of the solution (\ref{singularf}) into the equation (\ref{fmewode}) gives
\begin{equation}
\left( \delta ck^2 A p +\delta c k^2 A p^2 \right) \csch^{p+2}{\xi}+\left( \delta c k^2 A p^2-cA \right) \csch^{p}{\xi}+\epsilon k A^3  \csch^{3p}{\xi}=0
\end{equation}
Choice of $p=1$ gives
\begin{equation}
\left( A^3 \epsilon k +2Ac\delta k^2 \right) \csch^{3}{\xi}+\left( Ac\delta k^2 -Ac\right) \csch{\xi}=0 \label{eqq4}
\end{equation}
Solving the equation for nonzero $\csch{\xi}$ gives
\begin{equation}
\begin{aligned}
A&=\pm \sqrt{\pm 2\frac{c\sqrt{\delta}}{\epsilon}} \\
k&=\pm \frac{1}{\sqrt{\delta}} \label{solll}
\end{aligned}
\end{equation}
Thus, the singular solution of the FMEWE is formed as 
\begin{equation}
u(x,t)=A\csch \left ( \dfrac{\tilde{k}x^{\alpha}}{\Gamma{(1+\alpha)}}-\frac{\tilde{c}t^{\alpha}}{\Gamma{(1+\alpha)}}\right)
\end{equation}
where $A$ and $k$ are given in Eq.(\ref{solll}).
The singular solution simulations for various $\alpha$ values and $\epsilon =\delta =c=1$ with $k=1/\sqrt{\delta}$ and $A=\sqrt{2c\sqrt{\delta}/\epsilon}$ are plotted in Fig.(\ref{fig:4a}-\ref{fig:4d}).

\section{Conclusion}
  In the study, the bright soliton solutions and singular solutions for space-time fractional EW equation and modified EW equation are obtained by ansatz method. After reducing both equations to integer-ordered ordinary differential equations, the $\sech$ and $\csch$ type solutions are investigated. Substituting ansatzs into the integer-ordered ordinary differential equations, the nonzero coefficients in solutions are determined. The graphs of all solutions are depicted for suitable coefficients.

\end{document}